\documentstyle[aps,pre,epsf,multicol]{revtex}

\newcommand{\dir}{Figs/}

\begin{document}

\draft

\title{Molecular Dynamics Study of the Nematic-Isotropic Interface }

\author{
Nobuhiko Akino$^{1}$, Friederike Schmid$^{2}$, and 
Michael P. Allen$^{3}$
}

\address{ 
{$^1$ Max-Planck-Institut f\"ur Polymerforschung, Postfach 3148, D-55021 Mainz, Germany}\\
{$^2$ Fakult\"at f\"ur Physik, Universit\"at Bielefeld,
Universit\"atsstrasse 25, D-33615 Bielefeld, Germany}\\
{$^3$ H. H. Wills Physics Laboratory, University of Bristol,
Royal Fort, Tyndall Avenue, Bristol BS8 1TL, U.K.}\\
}

\date{November 7, 2000}

\maketitle

\begin{abstract}
We present large-scale molecular dynamics simulations of a
nematic-isotropic interface in a system of repulsive ellipsoidal 
molecules, focusing in particular on the capillary wave fluctuations of 
the interfacial position. The interface anchors the nematic phase
in a planar way, {\em i.e.}, the director aligns parallel to the
interface. Capillary waves in the direction parallel and perpendicular 
to the director are considered separately. We find that the spectrum 
is anisotropic, the amplitudes of capillary waves being larger in the 
direction perpendicular to the director. In the long wavelength limit, 
however, the spectrum becomes isotropic and compares well with the 
predictions of a simple capillary wave theory. 
\end{abstract}


\begin{multicols}{2}
 
\section{Introduction}

Surfaces and interfaces in liquid crystals have been the
subject of much interest both from a fundamental point of view and
because of their practical importance in the context of liquid crystal 
devices \cite{deg-prost,jerome}. 

Liquid crystals are formed by anisotropic particles. Depending on 
temperature and density, they exhibit various liquid phases.
Here, we shall be concerned with the isotropic phase (I), which 
is an ordinary fully symmetric fluid phase, and the nematic
phase (N), where the fluid retains the translational symmetry
in all directions (no positional order), but exhibits long range
orientational order. According to a very general symmetry argument,
the transition between the isotropic and the nematic phase is first 
order\cite{deg-prost}. Therefore there exist regions in phase space 
where nematic and isotropic fluids coexist and are separated by 
interfaces.

The nematic state is characterized by a so-called nematic director,
which specifies the preferred direction of alignment of the particles. 
Surfaces and interfaces break the isotropy of space and usually 
favour a certain director orientation. This effect, called surface 
{\it anchoring}, is commonly characterized by the tilt angle $\theta$
between the preferred orientation and the surface (interface) normal, 
and the anchoring strength or anchoring energy. In particular, the
cases $\theta = 0^\circ$ (perpendicular alignment) and $\theta = 90^\circ$ 
(parallel alignment) are often referred to as homeotropic anchoring and
planar anchoring, respectively.

Theoretically, particular attention has been given to NI-interfaces
in systems of hard particles. Studies of NI-interfaces in hard rod
systems based on Onsager's approach \cite{onsager} have indicated 
that the surface free energy has a minimum when the rods lie parallel 
to the interface\cite{chen-noolandi,koch-harlen}, suggesting that
the anchoring in this system is planar. Similar results have
been obtained in systems of hard ellipsoids\cite{allen}. 
According to the Onsager theory, the interfaces are flat, i.e., 
fluctuations of the interface position (capillary waves) are not present.

In general, however, one would expect such fluctuations in fluid-fluid 
interfaces. The usual simple argument reads as follows: 
On very large length scales, the free energy of a system containing an 
interface should be governed by the interfacial tension $\gamma$. 
Let us neglect bubbles and overhangs and parametrize the local position 
of a fluctuating planar interface by a single-valued function $h(x,y)$. 
Fluctuations enlarge the interfacial area and thus cost the free 
energy\cite{buff,cw}
\begin{equation}
\label{hcw}
{\cal F}_{\rm CW}\{ h \} = \frac{\gamma}{2} \int dxdy \left[ 
                                  \left( \frac{\partial h}{\partial x} 
                                   \right)^{2} 
                                 +\left( \frac{\partial h}{\partial y} 
                                   \right)^{2} 
                                  \right],
\end{equation}
(assuming small distortions with $|\partial h/\partial x|$,
$|\partial h/\partial y| \ll 1$). Since this free energy
functional is quadratic, the partition function can be evaluated 
exactly. ${\cal F} \{h\}$  is easily diagonalized by means of a 
two dimensional Fourier transform. One finds that the squared 
amplitude of fluctuations with wavevector ${\bf q}$ is on average
given by
\begin{equation}
\label{eq:hq}
\langle |h({\bf q})|^{2} \rangle = \frac{k_{\rm B} T}{\gamma q^{2}},
\end{equation}
i.e., {\em diverges} in the long-wavelength limit ${\bf q} \to 0$.
The argument thus not only predicts on fairly general grounds the existence 
of capillary waves at all temperatures, but also that they are actually 
quite large.

On the other hand, it has been argued that the Onsager approach should
be exact in the limit of infinitely elongated particles\cite{mao}. 
The Onsager theory assumes that the structure of the fluid is 
entirely determined by the second virial coefficient, {\em i.e.}, by the 
statistical properties of clusters of two particles. In systems of 
very long, very thin particles, the probability that one particle 
interacts with more than one other particle approaches zero, and the 
Onsager assumption becomes exact. One concludes that capillary waves
should vanish in this limit, in contradiction to the argument presented
just above.

\end{multicols}\twocolumn

How can this be? Two explanations are conceivable.
First, capillary waves vanish if the interfacial tension diverges. 
However, this is unlikely here, since the coexistence density of 
the nematic and the isotropic phase moves towards zero as the particle 
elongation is increased. Second, capillary waves might be suppressed by 
long-range interactions, which are disregarded in the functional (\ref{hcw}). 
Indeed, elastic long-range interactions are present in the nematic phase. 
We recall that the interface orients the director of the nematic 
liquid in a certain direction. Long-range interface fluctuations 
therefore lead to long-range elastic distortions of the director field, 
which are penalized.

In order to shed more light on these issues, we have carried out
extensive Molecular Dynamics simulations of an N-I interface in 
a system of over 100000 particles of elongation 15, and 
analyzed the capillary wave spectrum in detail.

To our knowledge, there exist only few numerical studies of
N-I interfaces \cite{bates-zannoni,allen,mc-allen-schmid}. 
In one of them \cite{bates-zannoni}, a system of molecules with 
length-to-width ratio $\kappa=3$ interacting via a Gay-Berne 
potential\cite{gay-berne} was studied in a simulation box with a 
temperature inhomogeneity, which served to make and maintain the NI
interface. 
The molecules in the nematic phase were found to align
parallel to the interface (planar anchoring). However, due to the use
of the temperature inhomogeneity, the system is not at thermodynamic
equilibrium. Capillary waves of the interface position fluctuations 
are obviously suppressed. In the other studies\cite{allen,mc-allen-schmid},
repulsive ellipsoids of revolution of axis ratio $\kappa=15$ were used in a 
system confined between two hard walls. The interactions of the walls 
with molecules were systematically varied to study the interplay between 
surface and interface anchoring. Planar alignment at the N-I interface 
was again observed. The system sizes considered were too small to
allow for an analysis of capillary waves. 

The present work builds on Reference \cite{mc-allen-schmid}. 
We study NI-interfaces in a very large system of repulsive ellipsoids
by means of extensive computer simulations, focusing in particular on 
the capillary wave fluctuations of the interfaces. The paper is
organized as follows: In the next section, we describe the model and
some simulation details . Section \ref{sec:gibbs} explains our 
way to determine the local position of the interface. The results are 
described and summarized in section \ref{sec:results}.

\section{Model and Simulation Details} \label{sec:model}

Our system consists of idealized ellipsoidal particles with elongation 
$\kappa=\sigma_{l}/\sigma_{s}=15$ where $\sigma_{l}$ and $\sigma_{s}$ 
are the length and width of the particles, respectively. The large 
value of the elongation $\kappa=15$ ensures that the order of the
transition is fairly strong, which makes it easier to generate
and maintain the interface. The interaction between two ellipsoids 
$i$ and $j$ is given by
\begin{equation}
\label{potential}
V({\bf u}_{i},{\bf u}_{j},{\bf r}_{ij})=
\left\{ \begin{array}{ll} 
 4\epsilon (X^{12}-X^{6}+\frac{1}{4}) & \mbox { $X^{6}>\frac{1}{2}$}  \\
     0                                & \mbox { otherwise} 
         \end{array}
\right.
\label{eq:potential}
\end{equation}
with
\begin{equation}
X=\frac{\sigma_{s}}
       {r-\sigma({\bf u}_{i},{\bf u}_{j},{\hat {\bf r}_{ij}})+\sigma_{s}}
\end{equation}
Here, ${\bf u}_{i}$ and ${\bf u}_{j}$ are the orientations of
ellipsoids $i$ and $j$, and ${\bf r}_{ij}$ is the center-center vector
between the ellipsoids, of magnitude $r$ and direction $\hat {\bf r}_{ij}$. 
The distance function $\sigma$ \cite{distance}
approximates the contact distance between two ellipsoids and is given by
\begin{eqnarray}
\sigma({\bf u}_{i},{\bf u}_{j},{\hat {\bf r}}) & = &
\sigma_{s} \Big\{ 1 - \frac{\chi}{2}
                         \Big[ \frac{ ({\bf u}_{i} \cdot {\hat {\bf r}}
                                  + {\bf u}_{j} \cdot {\hat {\bf r}})^{2} }
                                 {1+\chi {\bf u}_{i}\cdot{\bf u}_{j}}
\\
                   &&  \qquad + \; \frac{ ({\bf u}_{i} \cdot {\hat {\bf r}}
                                  - {\bf u}_{j} \cdot {\hat {\bf r}})^{2} }
                             {1-\chi {\bf u}_{i}\cdot{\bf u}_{j}}
                     \Big] \Big\} ^{-1/2},\nonumber
\end{eqnarray}
where the parameter $\chi$ is related to the elongation 
$\kappa \equiv \sigma_{l} / \sigma_{s}$ through
\begin{equation}
\chi=\frac{\kappa^{2}-1}{\kappa^{2}+1}.
\end{equation} 
Since the attractive tail of the potential eqn. (\ref{eq:potential}) has
been cut off, its range is much shorter than that of the Gay-Berne
potential. This enables us to simulate systems with a large number of 
particles, which is essential to study the capillary wave effects 
of interest to us.
For convenience, $\sigma_{s}=1$ defines a unit of length, $\epsilon=1$
a unit of energy, and we take particle mass $m=1$; the particle moment
of inertia is set at $I=50m\sigma_{s}^2$.

As a preliminary run, we performed a molecular dynamics (MD) simulation 
of a system with 7200 molecules at constant volume in a box geometry 
$(L_x:L_y:L_z) = (1:1:8)$ with periodic boundary conditions in all 
directions. The density was chosen in the coexistence region,
$\rho=0.017/\sigma_s^3$, and the temperature $k_{\rm B}T/\epsilon=1$. 
A nematic slab bounded by two interfaces parallel 
to the $xy$ plane assembled as a result. More than $5.5\times10^{6}$ MD 
steps were required to equilibrate this system, 
where each step covers 0.002 time units. 
The last configuration
was then reproduced $4$ times in the $x$ and $y$ direction, which
generated a system with $N=115200$ molecules and box size 
$L_{x}=L_{y}=150.1958 \sigma_s \equiv L$ and $L_{z}=300.3916 \sigma_s$.
This was used as the starting point for a MD simulation of $4.2\times10^{6}$
MD steps.

\section{Block Analysis and The Dividing Surface} \label{sec:gibbs}

In order to study the interfacial position fluctuations as a function
of the system size, one can either perform simulations of various system
sizes or simulate a single (large) system and analyze subsystems of
it. In this study, we took the latter approach. We split our system of 
size $L \times L \times L_{z}$ into columns of block size $B \times B$ 
\begin{figure}  
\centering\leavevmode
\epsfxsize 8.5cm
\epsfbox{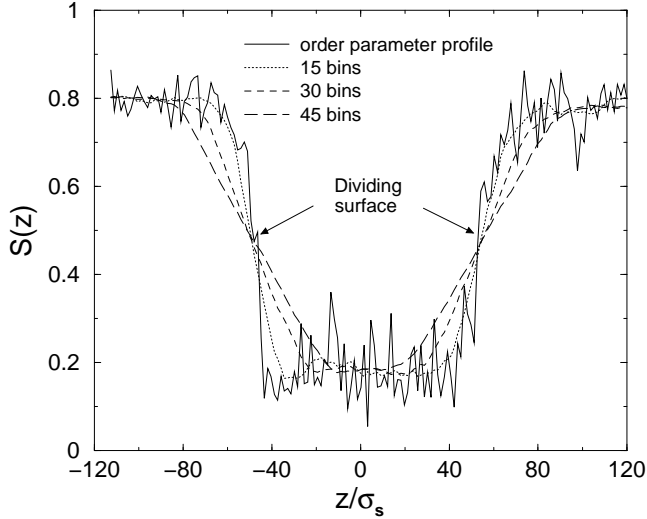}
\caption{Illustration of our method of estimating the dividing
  surface. Solid line shows an example of an order parameter profile
  for one of 16 columns with $B=L/4$. Dotted, dashed and long dashed
  lines show averages of this profile over $N_{\rm ave}=15$, $30$, and 45 
  bins, respectively. We defined the position of the interface $z_{\rm int}$ 
  to be the intersection of the two profiles averaged with $N_{\rm ave}=15$ 
  and $N_{\rm ave}=30$.} 
\label{fig:gibbs_method}
\end{figure}

\begin{figure}  
\centering\leavevmode
\epsfxsize 8.5cm
\epsfbox{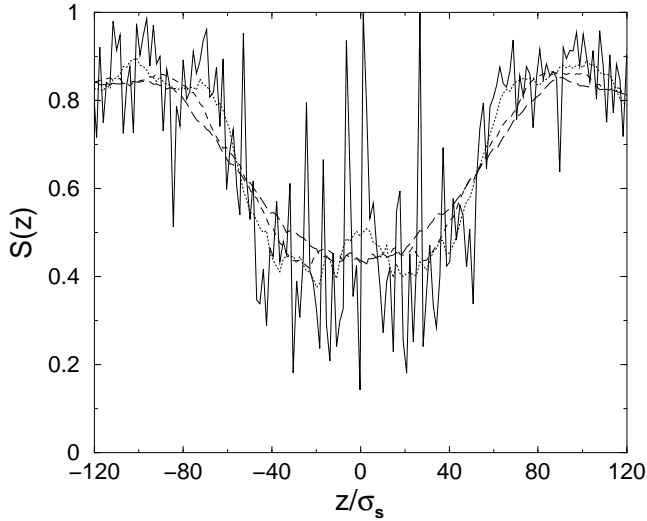}
\caption{Same as Fig.~1 but for $B=L/10$. 
The fluctuations in the order parameter profile are much larger.} 
\label{fig:gibbs_B10}
\end{figure}

\vspace*{0.5cm}

\noindent
and height $L_{z}$. The columns are further divided into $N_{\rm bins}=200$ 
bins in the $z$ direction. The local position of the interface 
$z_{\rm int}(x,y)$\cite{block} in each column is estimated as follows.

First, we compute the local ordering tensor ${\bf S}$ in each bin 
(of size $B \times B \times (L_{z}/N_{\rm bins})$), defined as 
\begin{equation}
S_{\alpha \beta}=\frac{1}{n}\sum_{i=1}^{n}
\frac{1}{2}(3u_{i\alpha}u_{i\beta}-\delta_{\alpha \beta}),
\label{eq:ordering}
\end{equation}
where $u_{i\alpha}$ is the $\alpha$ component of the unit vector which
points along the axis of the $i$-th molecule ($\alpha = x,y,z$),
and $n$ is the number of molecules in the bin.
The nematic order parameter $S(z)$ in the bin centered at $z$ is given by 
the largest eigenvalue of the ordering tensor. The further procedure
is illustrated in Fig.~\ref{fig:gibbs_method}.
Once the order parameter profile $S(z)$ is obtained (solid line), 
we compute at least two averaged profiles over $N_{\rm ave}$ bins, where 
$N_{\rm ave}$ is chosen such that an averaged profile still reflects the 
existence of the interface, but short range fluctuations
are averaged out (dotted and dashed lines in 
Fig.~\ref{fig:gibbs_method}). Finally, the position of the interface 
is estimated as the intersection of two averaged profiles.

The method is motivated by the following consideration:
If the order parameter $S(z)$ did not fluctuate at all in the bulk,
the intersection would locate a ``dividing surface'', where the 
negative order parameter excess on the nematic side just
balances the positive order parameter excess on the isotropic side.
It should then be independent of $N_{\rm ave}$, as long as 
$L \: N_{\rm ave}/N_{\rm bins}$ is larger than the interfacial
width and smaller than the total width of the nematic slab. 
This is not exactly true in an actual configuration due to
the bulk fluctuations of $S(z)$, but the methods still works
well even for small block sizes (see Fig.~\ref{fig:gibbs_B10}).

\section{Results} \label{sec:results}

The fluctuations of the total director (the direction of the
eigenvalue corresponding to the largest eigenvalue of the total
ordering tensor) turned out to be so slow that the director
hardly changed throughout the run. It always pointed in the
$y$ direction.  This made it convenient to resolve wave-vector components
along and perpendicular to the director, without the need to apply
any kind of director constraint.

We used the procedure sketched in section \ref{sec:gibbs} to
determine the local deviations $h(x,y)=z_{\rm int}(x,y)-{\bar z}_{\rm int}$ 
of the local interface position from its average for various
block sizes $B$. The landscape obtained for the block size $B=L/8$ was 
further analyzed and Fourier transformed. Fourier modes $h({\bf q})$ are 
labeled by ${\bf n}=(n_{x},n_{y})$, where $n_x$ and $n_y$ are positive 
integers with ${\bf q}=\frac{2\pi}{L}{\bf n}$.

\begin{figure}
\centering\leavevmode
\epsfxsize 8.5cm
\epsfbox{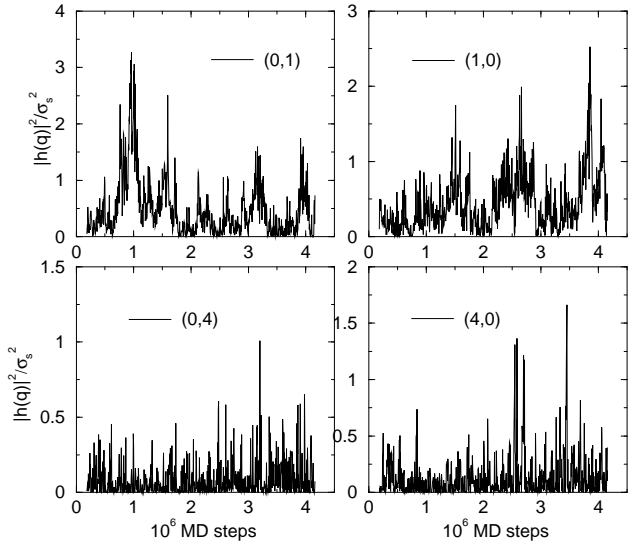}
\caption{Fourier modes $|h({\bf q})|^2$ for ${\bf n}=(0,1)$, $(1,0)$, $(0,4)$,
  and $(4,0)$ vs. time in units of MD steps.} 
\label{fig:time_evol}
\end{figure}

First, we inspect the relaxation times and the correlation times
of the Fourier modes. The time evolution of the modes
${\bf n}=(0,1)$, $(1,0)$, $(0,4)$, and $(4,0)$, is shown as a 
function of time (in MD steps) in Fig.~\ref{fig:time_evol}.
From there we estimate the number of MD steps needed to equilibrate
the system. For the slowest mode ${\bf n}=(0,1)$, in the direction
parallel to the director, the equilibration process seems to require
roughly $1.0\times10^{6}$ MD steps. Based on this information, we   
have discarded the initial $1.2 \times 10^{6}$ MD steps, and
collected results over the following $2.96\times10^{6}$ MD steps only.

One also has to ensure that the total length of our simulation run 
significantly exceeds the characteristic time 

\begin{figure}
\centering\leavevmode
\epsfxsize 8.5cm
\epsfbox{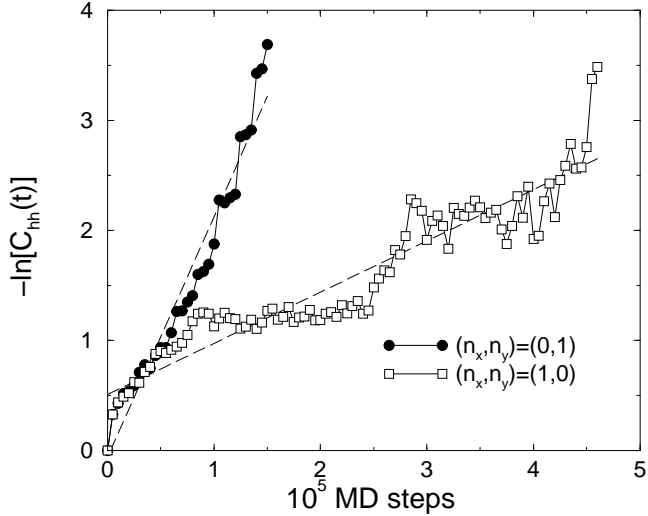}
\caption{Autocorrelation function $C_{hh}$ (log scale) vs. MD steps for the
two slowest modes  ${\bf n}=(1,0)$ (normal to the director) and 
${\bf n}=(0,1)$ (parallel to the director).  Dashed lines are linear
fits to give the correlation time.}
\label{fig:auto_corr}
\end{figure}

\vspace*{0.5cm}

\noindent 
scale of the slowest
capillary wave mode. In order to check this, we have computed the 
autocorrelation functions of the squared amplitude of the Fourier
modes $h({\bf q})$ 
\begin{equation}
C_{hh}(t)=\frac{\langle |h(t)|^{2}|h(0)|^{2} \rangle - 
\langle |h|^{2} \rangle^{2}}
               {\langle |h|^{4} \rangle - \langle |h|^{2} \rangle^{2}}
\propto e^{-t/\tau}
\end{equation}

They are shown in Fig.~\ref{fig:auto_corr} for the slowest Fourier 
modes in the direction normal to the director ${\bf n}=(1,0)$ and 
parallel to the director ${\bf n}=(0,1)$.
The correlation time is estimated from the slope of the fitted line
as $\tau \approx 1.0\times10^{5}$ MD steps for the $(1,0)$-mode and 
$\tau \approx 3.0\times10^{5}$ MD steps for the $(0,1)$-mode. 
This is only by a factor of 10 smaller 

\begin{figure}  
\centering\leavevmode
\epsfxsize 8.5cm
\epsfbox{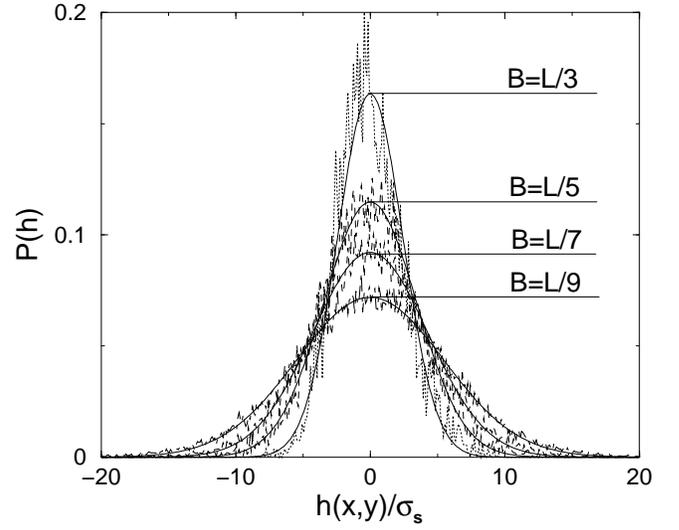}
\caption{Distribution of local interface positions  
for block sizes $B=L/3$, $L/5$, $L/7$  and $L/9$.
The solid lines are Gaussian fits (eq.(9).}
\label{fig:distribution}
\end{figure}

\vspace*{0.5cm}

\noindent
than the length of the
total simulation run. Hence the statistical error of our results 
is quite large.

We can now proceed to compare our results with the predictions of
the capillary wave theory. 

First, we consider the distribution of $h(x,y)$. 
From the free energy functional (\ref{hcw}), one derives 
the exact prediction \cite{weeks}

\begin{equation}
P(h)=\frac{1}{\sqrt{2\pi s^{2}}} \exp \left( -\frac{h^{2}}{2s^{2}}
\right)
\label{eq:gaussian},
\end{equation}
where 
\begin{eqnarray}
\label{eq-s2}
s^{2} &=& \langle h^{2}(x,y) \rangle \nonumber \\
      &=& \frac{1}{4\pi^{2}} \int d{\bf q} \langle |h({\bf q})|^{2}
                                           \rangle
       = \frac{1}{2\pi\gamma} \ln \left( \frac{q_{\rm max}}{q_{\rm min}} \right).
\end{eqnarray}
The lower and upper cutoffs $q_{\rm max}=2 \pi/B$ and $q_{\rm min}=2 \pi/L$
come into play, because the integral 
$\int d{\bf q} \langle |h({\bf q})|^{2}\rangle\sim\int dq/q$ 
diverges as $q \rightarrow 0$ and $q \rightarrow \infty$.
The actual local height distribution $P(h)$ obtained from our
simulations is plotted for several block sizes in 
Fig.~\ref{fig:distribution}. It can be fitted nicely by
a Gaussian distribution (\ref{eq:gaussian}).

Next, we study the width $\omega$ of the average order parameter 
profile as a function of the block size. According to the capillary
wave theory, the capillary wave fluctuations broaden it for
large blocks $B$ according to
\begin{equation}
\omega^{2}=\omega_{0}^{2}+\frac{\pi}{2}s^{2}
          =\omega_{0}^{2}+\frac{k_{\rm B} T}{4\gamma}
           \ln \left( \frac{\hat{q}_{\rm max}}{\hat{q}_{\rm min}} \right).
\label{eq:width}
\end{equation}
Here the cutoff wavevectors are given by $\hat{q}_{\rm min} = 2 \pi/B$ and 
$\hat{q}_{\rm max}=2 \pi/a_0$, where $a_0$ is a microscopic length which 
need not be specified here.

\begin{figure}  
\centering\leavevmode
\epsfxsize 8.5cm
\epsfbox{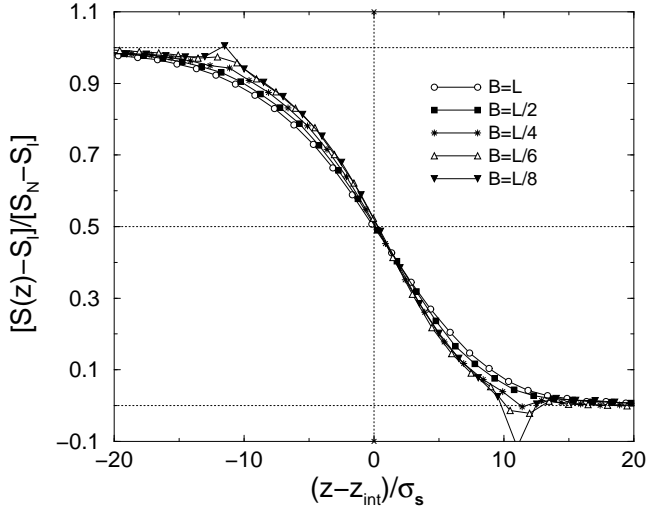}
\caption{Order parameter profiles for block sizes $B=L$, $L/2$,
 $L/4$, $L/6$ and $L/8$. $S_{\rm N}$ and $S_{\rm I}$ are the average values
 of the order parameter in the nematic and isotropic phase,
 respectively.} 
\label{fig:profile}
\end{figure}

\begin{figure}  
\centering\leavevmode
\epsfxsize 8.5cm
\epsfbox{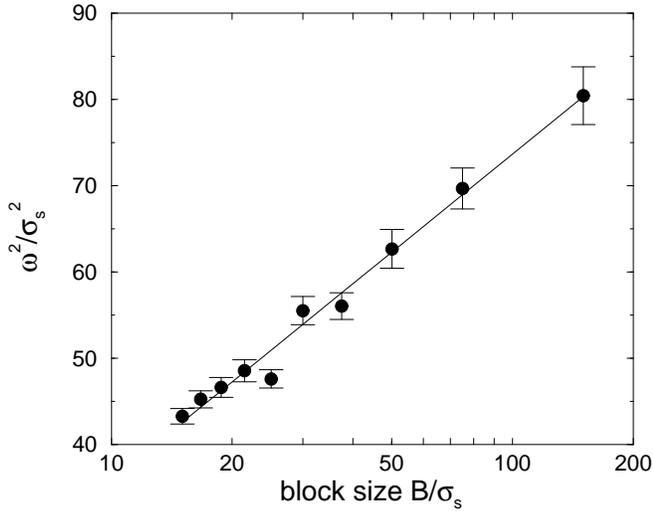}
\caption{Squared interfacial width $\omega^{2}$ vs. block size
  $B$. Solid line is a fit to eq.(11). }
\label{fig:w2vsb}
\end{figure}

\vspace*{0.5cm}
The broadening effect is demonstrated for different block sizes $B$ in 
Fig.~\ref{fig:profile}. The interfacial width can be estimated 
by fitting order parameter profiles such as those shown in 
Fig.~\ref{fig:profile} to {\it tanh}-profiles, 

\begin{equation}
S(z)=\frac{1}{2} \left( S_{\rm N} + S_{\rm I} \right)
    +\frac{1}{2} \left( S_{\rm N} - S_{\rm I} \right) 
                        \tanh \left( \frac{z-z_{\rm int}}{\omega} \right),
\label{eq:mean-field}
\end{equation}
where the parameters $S_{\rm N}$ and $S_{\rm I}$ are the values of the order 
parameter in the bulk nematic and isotopic phases. (Note that $S_{\rm I}$ is 
nonzero due to the finite number of particles in a bin).
Fig.~\ref{fig:w2vsb} plots the squared interfacial width $\omega^{2}$ 
as a function of block size $B$. One clearly observes the
logarithmic increase of $\omega^{2}$ with block size $B$
predicted by eqn. (\ref{eq:width}). From the slope of the line
one can estimate the interfacial tension, 
$\gamma = 0.016 \pm 0.002 k_{\rm B} T/\sigma_s^2$.

For comparison, we have also determined the interfacial tension from
the anisotropy of the pressure tensor, making use of
the relation \cite{schofield,mc-allen-schmid}
\begin{equation}
\label{schofield}
\gamma = \int_{-\infty}^{\infty} dz \: [P_{\rm N}(z)-P_{\rm T}(z)],
\end{equation}
where $P_{\rm N}$ and $P_{\rm T}$ are the normal and transverse pressure tensor
components. Here, we consider particles with pair interactions 
$V_{ij}$, eqn~(\ref{potential}) and systems with two interfaces in the 
$(xy)$-plane. Neglecting finite size effects and interactions between 
the interfaces, eqn. (\ref{schofield}) thus reads 
\begin{equation}
\gamma
= \frac{1}{4 L^2} \sum_{i<j} \left\langle 
r_{ij}^{x} \frac{\partial V_{ij}}{\partial r_{ij}^{x}}
+ r_{ij}^{y} \frac{\partial V_{ij}}{\partial r_{ij}^{y}}
- 2 r_{ij}^{z} \frac{\partial V_{ij}}{\partial r_{ij}^{z}}
\right\rangle \:.
\end{equation}
The simulation data yield $\gamma=0.0093 \pm 0.003 \sigma_s^2/k_{\rm B} T$.
This value agrees with earlier results obtained in smaller 
systems\cite{footnote}. It is of the same order, but smaller
than the value estimated from the interfacial broadening.
The quantitative difference possibly stems from attractive interactions 
between the two interfaces of the nematic slab. More simulations, 
in which the thickness of the slab is varied systematically, would be 
needed to elucidate this point.

\begin{figure}  
\centering\leavevmode
\epsfxsize 8.5cm
\epsfbox{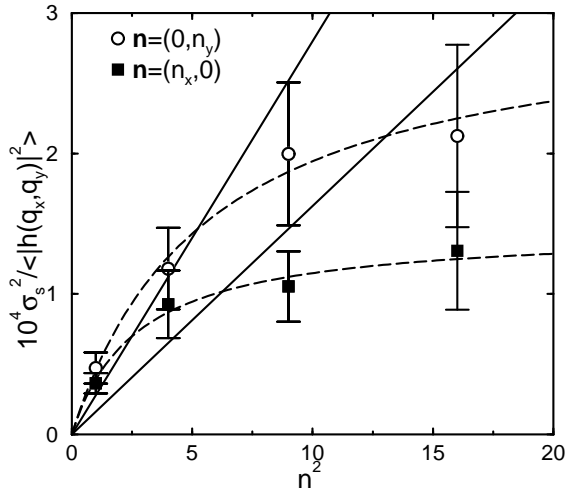}
\caption{Inverse of the mean-squared Fourier components of 
the interface position $1/\langle |h({\bf q})|^{2} \rangle$
vs square of the wave-vector, $n^2=(qL/2\pi)^{2}$ 
for ${\bf q}$ parallel (circle) and normal (square) to the
director. Dashed curves are guides to eye. Solid lines correspond to
the prediction of the simple capillary wave theory (eqn (2))
with $\gamma=0.016\sigma_{s}^{2}/k_{B}T$ (upper line), and
$\gamma=0.0093\sigma_{s}^{2}/k_{B}T$ (lower line).
}
\label{fig:h2_vs_q2}
\end{figure}

Finally, we turn to the analysis of the capillary wave spectrum
$\langle | h({\bf q}) |^2 \rangle$, which is predicted to be 
inversely proportional to $q^2$ according to eqn~(\ref{eq:hq}).
We study separately the ${\bf q}$-direction parallel 
(${\bf q} \propto (0,1)$) and perpendicular (${\bf q} \propto (1,0)$)
to the director. The inverse of $\langle | h({\bf q}) |^2 \rangle$
in these two directions is shown in Fig.~\ref{fig:h2_vs_q2}
as a function of the squared wave vectors. In the long wavelength 
limit ${\bf q}\rightarrow 0$, the spectrum appears to be isotropic,
and $1/\langle | h({\bf q}) |^2 \rangle$ approaches zero in
agreement with the capillary wave theory. The initial slope
is consistent with eqn. (\ref{eq:hq}) if one uses 
$\gamma = 0.016 k_{\rm B} T/\sigma_s^2$, and still roughly 
compatible if one uses $\gamma = 0.0093 k_{\rm B} T/\sigma_s^2$.
As ${\bf q}$ increases, the capillary wave spectrum becomes 
anisotropic. As one might expect intuitively, the amplitudes 
$\langle | h({\bf q}) |^2 \rangle$ are larger for capillary waves 
in the direction perpendicular to the director. 

Deviations of capillary wave spectra from the straight line
predicted by the simple capillary wave theory (\ref{eq:hq}) are 
often discussed in terms of higher order terms in the capillary
wave Hamiltonian (\ref{hcw}), {\em i.e.}, terms proportional
to squares of higher order derivatives of $h(x,y)$. 
For example, including the next-to-leading term leads to a 
prediction of the form $1/\langle | h({\bf q}) |^2 \rangle 
\propto \gamma q^2 + \delta q^4$, where $\gamma$ is the 
interfacial tension and $\delta$ is a bending rigidity. 
One might intuitively expect that the NI-interface has
bending stiffness, based on the argument that the elastic 
interactions should penalize interfacial bending. 

However, Fig.~\ref{fig:h2_vs_q2} indicates that the sign 
of the ``bending energy'' is {\em negative}. The naive
argument sketched above 

\vspace*{0.5cm}

\noindent
is thus clearly wrong. On the contrary, 
our unexpected result suggests that the elastic interactions 
influence the capillary waves on the largest length scale:
The interface is rougher on short length scales than
one would expect from the long wavelength fluctuations.
This observation is consistent with our speculations in
the introduction, that the elastic interactions may be
responsible for the suppression of capillary wave fluctuations 
of the NI-interface in the Onsager-limit of infinitely elongated particles.

We note that this is not the first observation of a negative
bending rigidity at a fluid-fluid interface. A similar
phenomenon has been predicted theoretically for liquid-vapour
interfaces \cite{blokhuis1,mecke} and verified
experimentally by C. Fradin {\em et al} \cite{fradin}. The negative 
bending rigidity is attributed in their case to long range
van der Waals interactions\cite{napiorkowsky}. Indications of a negative 
bending rigidity at a liquid-liquid interface have also been found in 
Molecular dynamics simulations of simple liquids by Stecki and 
Toxvaerd\cite{stecki,blokhuis2}.

To summarize, we have studied the NI-interface by large scale molecular 
dynamics simulation of a system of repulsive ellipsoidal molecules.
We find that the standard capillary wave theory explains most of
our results very well. Discrepancies are encountered when looking
at the amplitudes of capillary wave modes at large wave vectors,
{\em i.e.}, on small length scales. We find that they are smaller
than expected and anisotropic.

\acknowledgements

The simulations have been performed on the CRAY T3E of the
HLRZ in J\"ulich. N.A. received financial support from the
German Science Foundation (DFG).  MPA acknowledges the support of 
the Alexander von Humboldt Foundation and the Leverhulme Trust.
The parallel MD program used in this work was originally developed by
the EPSRC Complex Fluids Consortium.


\vspace*{-0.4cm}

\begin{thebibliography}{99}
\bibitem{deg-prost}
P.~G.~de~Gennes and J.~Prost, {\it The Physics of Liquid Crystals},
Clarendon Press, Oxford, 2nd edn., 1993.
\bibitem{jerome}
B.~Jerome, Rep. Prog. Phys. {\bf 54}, 391 (1991).
\bibitem{onsager}
L.~Onsager, Ann. N.Y. Acad. Sci. {\bf 51}, 627 (1949).
\bibitem{chen-noolandi}
Z.~Y.~Chen and J.~Noolandi, Phys. Rev. A{\bf 45}, 2389 (1992).
\bibitem{koch-harlen}
D.~L.~Koch and O.~G.~Harlen, Macromolecules {\bf 32}, 219 (1999).
\bibitem{allen}
M.~P.~Allen, J. Chem. Phys. {\bf 112}, 5447 (2000).
\bibitem{buff}   
  F. P. Buff, R. A. Lovett, F. H. Stillinger, 
  Phys. Rev. Lett. {\bf 15}, 621 (1965).
\bibitem{cw}
J.~S.~Rowlinson and B.~Widom, Molecular Theory of Capillarity
(Clarendon, Oxford, 1982).
\bibitem{mao}
Y.~Mao, M.~E.~Cates, H.~N.~W.~Lekkerkerker, J. Chem. Phys. 
{\bf 106}, 3721 (1997).
\bibitem{bates-zannoni}
M.~A.~Bates and C.~Zannoni, Chem. Phys. Lett. {\bf 280}, 40 (1997).
\bibitem{mc-allen-schmid}
A.~J.~McDonald, M.~P.~Allen, and F.~Schmid, cond-matt/0008056,
Phys. Rev. E (to appear).
\bibitem{gay-berne}
J.~G.~Gay and B.~J.~Berne, J. Chem. Phys. {\bf 74}, 3316 (1981).
\bibitem{distance}
B.~J.~Berne and P.~Pechukas, J. Chem. Phys. {\bf 56}, 4213 (1975).
\bibitem{block}
A.~Werner, F.~Schmid, M.~M\"uller, and K.~Binder, Phys. Rev. E {\bf
  59}, 728 (1999).
\bibitem{weeks}
  J. D. Weeks, J. Chem. Phys. {\bf 67}, 3106 (1977);
  D. Bedeaux and J. D. Weeks, J. Chem. Phys. {\bf 82}, 972 (1985).
\bibitem{schofield}
  P.~Schofield and J.~R.~Henderson,
  Proc. Roy. Soc. London A {\bf 379}, 231 (1982).
\bibitem{footnote} The simulations of ref. \cite{mc-allen-schmid}
  yielded the interfacial tension $\gamma = 0.011 \sigma_s^2/k_B T$ 
  However, the systems studied there were so small that the actual 
  value of $\gamma$ is not very reliable.
\bibitem{blokhuis1} E. M. Blokhuis, D. Bedeaux,
  Mol. Phys. {\bf 80}, 705 (1993). 
\bibitem{napiorkowsky} For a more general discussion of bending rigidities
  in isotropic fluids with arbitrary pair interactions see also
  N.~Napiorkowsky, S.~Dietrich,
  Phys. Rev. E {\bf 47}, 1836 (1993);
  Z. Phys. B {\bf 97}, 511 (1995).
\bibitem{mecke}
  K.~R.~Mecke and S.~Dietrich, 
  Phys. Rev. E {\bf 59}, 6766 (1999). 
\bibitem{fradin}
  C.~Fradin, D.~Luzet, D.~Smilgies, A.~Braslau, M.~Alba, N.~Boudet, 
  K.~Mecke, and J.~Daillant, Nature {\bf 403}, 871 (2000).
\bibitem{stecki} J. Stecki, S. Toxvaerd, 
  J. Chem. Phys. {\bf 103}, 9763 (1995). 
\bibitem{blokhuis2} E. M. Blokhuis, D. Bedeaux,
  Mol. Phys. {\bf 96}, 397 (1999).
\end{thebibliography}
\end{document}